\documentclass[aps,prb,twocolumn,superscriptaddress,showpacs]{revtex4}
\usepackage[T1]{fontenc}
\usepackage[latin9]{inputenc}
\usepackage{amsmath}
\usepackage{amssymb}
\usepackage{graphicx}
\usepackage{bm}
\usepackage{amsfonts,color}

\renewcommand{\vec}[1]{\bm{#1}}

\DeclareMathAlphabet{\mathsfsl}{OT1}{cmss}{m}{sl}

\begin{document}

\title{Numerical study of the negative nonlocal resistance and the backflow current
in a ballistic graphene system}

\author{Zibo Wang}
\affiliation{Microsystem and Terahertz Research Center, China Academy of Engineering Physics,
Chengdu 610200, China}
\affiliation{Institute of Electronic Engineering, China Academy of Engineering Physics, Mianyang 621999, China}

\author{Haiwen Liu}
\affiliation{Center for Advanced Quantum Studies, Department of Physics, Beijing Normal University, Beijing 100875, China}

\author{Hua Jiang}
\thanks{\texttt{jianghuaphy@suda.edu.cn}}
\affiliation{School of Physical Science and Technology, Soochow University, Suzhou 215006, China}
\affiliation{Institute for Advanced Study of Soochow University, Suzhou 215006, China}

\author{X. C. Xie}
\affiliation{International Center for Quantum Materials, School of Physics, Peking University, Beijing 100871, China}
\affiliation{CAS Center for Excellence in Topological Quantum Computation,
University of Chinese Academy of Sciences, Beijing 100190, China}
\affiliation{Beijing Academy of Quantum Information Sciences, West Bld.3,
No.10 Xibeiwang East Rd., Haidian District, Beijing 100193, China}

\date{\today}

\begin{abstract}
Besides the giant peak of the nonlocal resistance $R_{NL}$,
an anomalous negative value of $R_{NL}$ has been observed in graphene systems,
while its formation mechanism is not quite understood yet.
In this work, utilizing the non-equilibrium Green's function method,
we calculate the local-current flow in an H-shaped non-interacting graphene system
locating in the ballistic regime.
Similar to the previous conclusions made from the viscous hydrodynamics,
the numerical results show that
a local-current vortex appears between the nonlocal measuring terminals,
which induces a backflow current and a remarkable negative voltage drop at the probe.
Specifically, the stronger the vortex exhibits, the more negative $R_{NL}$ manifests.
Besides, a spin-orbital coupling is added as an additional tool to study this exotic vortex,
which is not a driving force for the arising vortex at all.
Moreover, a breakdown of the nonlocal Wiedemann-Franz law is obtained in this ballistic system,
and two experimental criteria are further provided to confirm the existence of this exotic vortex.
Notably, a discussion is made that the vortex actually originates from
the collision between the flowing current and the boundaries,
due to the long electron mean free path and the consequent ballistic transport
caused by the specific linear spectrum of graphene.
\end{abstract}

\date{\today}

\maketitle

\section{Introduction}

Nonlocal measurement indicates the detection of a voltage signal away from the path
that the current is expected to follow,
and has been developed as a powerful tool to
discover nontrivial interactions difficult to detect directly,
such as the electron-electron ($ee$) interaction, viscosity,
spin-orbital coupling, etc
\cite{Huber,Abanin1,Balakrishnan,Nowack,Gorbachev,Shimazaki,Sui,Michihisa,Bandurin1,Bandurin2,Braem,Wu,Berdyugin}.
Besides the giant peak of the nonlocal resistance $R_{NL}$,
more and more experiments have observed an anomalous negative value of the nonlocal resistance
\cite{Bandurin1,Bandurin2,Berdyugin},
which means the local current close to the nonlocal terminals
flows in direction opposite to the injected current.
Till now, kinds of theories have been proposed to explain this exotic phenomenon,
such as the viscous hydrodynamic fluid\cite{Bandurin1,Levitov,Alekseev,Levin},
the ballistic transport\cite{Wang,Tuan,Shytov,Chandra},
the magnetoelectric coupling\cite{Huang} and so on.

Among these theories, viscous hydrodynamics is the most developed one
\cite{Bandurin1,Levitov,Alekseev,Levin},
and has been supported by the recent experiments
for a two-dimensional fluid of electrons in graphene\cite{Berdyugin,Gallagher}.
The condition for viscous hydrodynamics happens when the extrinsic scattering length $r$
is far greater than the $ee$ collision mean free path $l_{ee}$.
That is to say, electrons injected into the system
must undergo frequent collisions with each other inside the bulk,
which can be well described by the theory of hydrodynamics
\cite{Jong,Scaffidi,Lucas}.
For instance, Bandurin et al. detected an anomalous nonlocal negative voltage drop
in a multi-terminal graphene device on the order of $1\mu m$\cite{Bandurin1}.
By solving the hydrodynamic equations,
a submicrometer-size whirlpool in the electron flow can be obtained.
Moreover, the calculation results show that the whirlpool enhances with the increase of the viscosity,
and disappears with zero viscosity, which fits the experiment well.
Thus, this negative nonlocal resistance is attributed to the formation of a local-current vortex.

More comprehensively,
some literatures further propose that in the $ee$ interaction dominated
but quasi-ballistic regime,
where $r$ is much smaller than $l_{ee}$,
the negative value of the nonlocal resistance can occur as well\cite{Shytov,Bandurin2,Braem}.
Furthermore, as a direct prove to the ballistic transport without $ee$ collision,
our previous work and others
based on the non-equilibrium Green's function (NEGF) method
predict that the ballistic transport itself could induce a negative nonlocal resistance
\cite{Wang,Tuan}.
In addition, with the help of the semi-classical Boltzmann equation and phenomenological conditions,
recent work could even obtain a vortex in the ballistic regime classically\cite{Chandra}.
However, compared with the well-studied theory in hydrodynamics,
the formation mechanism of the negative $R_{NL}$ observed in ballistic system
without $ee$ interaction desperately needs in-depth analysis,
such as the demonstration of the local-current flow in the view of quantum transport.

In this work, we further study the origin of the negative nonlocal resistance
in a non-interacting graphene model with an external Rashba effect,
and try to confirm whether the local-current vortex
also exists in this ballistic system, similar as that in a viscous fluid\cite{Bandurin1}.
First, in an H-shaped four-terminal graphene system,
we calculate the local-current distribution with
different Fermi energy $E_F$ and different Rashba strength $\lambda_R$.
The results show that an exotic vortex emerges between the two nonlocal measuring terminals
similar to that in a viscous fluid.
Further calculation indicates that this vortex strengthens with the decrease of $\lambda_R$.
In comparison with the fact that the negative value of the nonlocal resistance
also increases with weakening Rashba effect,
we confirm that the vortex has a positive relation to the negative nonlocal resistance.
Thus, a conclusion can be made that the ballistic transport in the nonlocal measurement
does exist in the form of a vortex,
which consists of backflow current and induces a negative contribution to the nonlocal resistance.
Then, the local and nonlocal thermal conductance of this system are also obtained within the NEGF framework.
In comparison to the scaled local (nonlocal) electrical conductance
$\mathcal{L}_0^{-1}R_L/T$ ($\mathcal{L}_0^{-1}R_{NL}/T$),
we find that our ballistic system actually breaks the nonlocal Wiedemann-Franz (WF) law
instead of the local one found in hydrodynamics\cite{Crossno},
and this violation gradually disappears with weakening vortex strength.
Based on the above results, we further propose two experimental methods
to confirm the existence of the local-current vortex in a ballistic system.
Specifically, with increasing $\lambda_R$,
both the negative value of $R_{NL}$ and the breakdown of the nonlocal WF law
should gradually disappear in ballistic regime.
At last, a discussion is made that it is actually
the collision between the flowing current and the boundaries
that induces this exotic vortex.
Since this collision highly relies on a long electron mean free path $l_e$
and the consequent ballistic transport,
the vortex will be much easier to be observed in graphene
due to its unique linear energy dispersion.

The rest of this paper is organized as follows.
In Sec.II, we first numerically obtain the local-current flow
and find a vortex in an H-shaped four-terminal ballistic graphene system.
Then, in Sec.III, we propose two experimentally feasible methods
to confirm the existence of this vortex indirectly in the ballistic regime.
Next, a discussion about the possible origin of the vortex flow is made in Sec.IV.
And finally, a conclusion is presented in Sec.V.

\section{Local-current flow in H-shaped four-terminal system}

\begin{figure}[h]
\includegraphics [width=\columnwidth, viewport=0 0 815 476, clip]{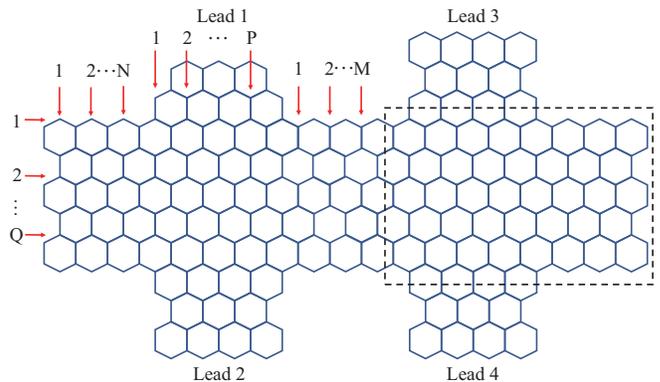}
\caption{
The schematic diagram of the H-shaped four-terminal system.
The current is injected from lead 1 to lead 2.
The voltage signal is detected on lead 12 for the local resistance,
and on lead 34 for the nonlocal resistance.
The rectangle circled by the black dashed line indicates the region
where the local-current flow is calculated.
}\label{bulkbandsevolution}
\end{figure}

In this work, we consider an H-shaped four-terminal graphene system\cite{foot1},
whose schematic diagram is shown in Fig.1(a),
to simulate the local current distribution in the nonlocal measuring experiment.
The central region for this system is marked with parameters $M,N,P$ and $Q$.
For instance, in Fig.1, we show a sample with $M=N=Q=3$ and $P=4$.
During the calculation process,
the current is injected from lead 1 and flows out from lead 2.
Meanwhile, the nonlocal voltage is detected between lead 3 and 4.
In this paper, we only show the local current distributed
in the rectangle region, which is surrounded by the black dashed line.
This is due to the fact that we find the local-current flow exhibits nontrivial characteristics
only between the nonlocal measuring terminals,
and exhibits the classical Ohmic distribution in other regions.

Similar as our previous work of Ref.[\onlinecite{Wang}],
the tight-binding Hamiltonian for this system is written as:
\begin{eqnarray}
H=\sum_i\epsilon_ic_i^\dagger c_i-t\sum_{\langle ij\rangle} c_i^\dagger c_j+i\lambda_R\sum_{\langle ij\rangle}c_i^\dagger(\vec{s}\times\hat{\vec{d}}_{ij})_zc_j.
\end{eqnarray}
The first term is the on-site potential with $\epsilon_i$ on the $i$th carbon atom.
The second term represents the nearest-neighbor hopping with strength $t$.
In the following calculation, we take $t$ as the energy unit
and all other parameters are normalized based on $t=2.8$eV.
The last term describes the external Rashba effect with strength $\lambda_R$.
The disorder existing in the central region is modeled by Anderson disorder
with random potential uniformly distributed in $[-w/2,w/2]$,
where $w$ is the disorder strength.
In this work, we choose $w=1$.

\begin{figure*}[htbp!]
\scalebox{2}{\includegraphics [width=\columnwidth, viewport=0 0 1001 433, clip]{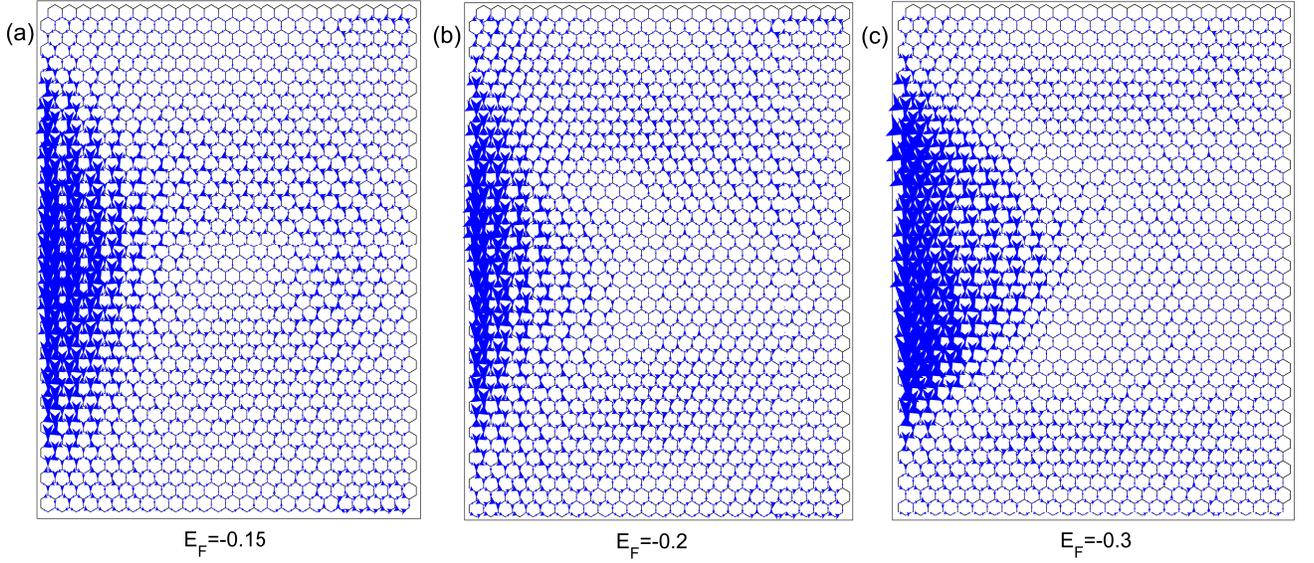}}
\caption{
The local current flow in the black dashed rectangle marked in Fig.1 when $\lambda_R=0.1$.
(a) $E_F=-0.15$; (b) $E_F=-0.2$; (c) $E_F=-0.3$.
The size of each arrow is proportional to the strength of the local current vector.
It is obvious that a counter-clockwise vortex emerges in panel (a)
and gradually disappears with the increase of $|E_F|$ in panels (b) and (c).
All three panels have sufficiently high resolutions to be zoomed in to show the vortex clearly.
}\label{bulkbandsevolution}
\end{figure*}

In Ref.[\onlinecite{Wang}],
we have specifically introduced how to calculate the local and the nonlocal resistance:
$R_L=(V_1-V_2)/I_1$ and $R_{NL}=(V_3-V_4)/I_1$ based on the Landauer-Buttiker formula\cite{book}.
Since it's the local-current distribution that we focus on in this work,
the size of the central region is limited
for the convenience and clearness to read the local-current flow vectors inside the black dashed line.
Thus, the size parameters $M$, $N$, $P$ and $Q$ are chosen as $M=10$, $N=5$, $P=20$ and $Q=20$,
which means the size of this system we calculated is about $20nm\times10nm$\cite{foot2}.

In order to simulate the local-current flow in the nonlocal measuring experiment,
a small external voltage bias $V=V_1-V_2$ is applied between lead 1 and 2.
With the help of the NEGF method,
the local current flows from site $i$ to its neighbor $j$ can be deduced as\cite{Jauho,Jiang}:
\begin{eqnarray}
J_{i\rightarrow j} = -\frac{2e}{h}\sum_{\alpha,\beta}\int^{\infty}_{-\infty}{\rm d}E
{\rm Re}[H_{i\alpha,j\beta}G^K_{j\beta,i\alpha}(E)],
\end{eqnarray}
where $\alpha$,$\beta$ denote the spin indices,
and $G^K_{j\beta,i\alpha}(E)$ is the Keldysh Green's function.
When the applied voltage $V_{1,2}$ is small and the system is in the zero temperature,
by applying the Keldysh equation $G^K=G^r\sum_n i\Gamma_n f_n G^a$ and assuming $V_1>V_2$,
the Eq.(2) can be rewritten as:
\begin{eqnarray}
J_{i\rightarrow j}
&=& \frac{2e}{h}\sum_{\alpha,\beta}\int^{eV_2}_{-\infty}{\rm d}E
{\rm Im}\{H_{i\alpha,j\beta}[G^r\sum_n \Gamma_nG^a]_{j\beta,i\alpha}\} \nonumber \\
&+& \frac{2e^2}{h}\sum_{\alpha,\beta}{\rm Im}
[H_{i\alpha,j\beta}\sum_nG^n_{j\beta,i\alpha}(E_F)(V_n-V_2)],
\end{eqnarray}
where $V_n$ is the voltage of the $n$th lead,
and can be obtained in the calculation of $R_{NL}$.
$G^n(E)=G^r(E)\Gamma_n(E)G^a(E)$ is the electron correlation function.
The linewidth function is
$\Gamma_n(E)=i\{\Sigma^r_n(E)-[\Sigma^r_n(E)]^{\dagger}\}$,
and the Green's function reads $G^r(E)=[G^a(E)]^{\dagger}
=[EI-H_{cen}-\sum_n\Sigma^r_n(E)]$.
Here, $\Sigma^r_n(E)$ is the retarded self-energy
due to the coupling to the $n$th lead,
and $H_{cen}$ is the Hamiltonian used in the central region.
The first part of Eq.(3) gives rise to the equilibrium current $\vec{J}_{eq}$,
which equals zero due to the time-reversal symmetry of the Hamiltonian described by Eq.(1).
Thus, the local current flows from site $i$ to its neighbor $j$ can be simplified as:
\begin{eqnarray}
J_{i\rightarrow j}
=\frac{2e^2}{h}{\rm Im}\sum_{\alpha,\beta}
[H_{i\alpha,j\beta}\sum_nG^n_{j\beta,i\alpha}(E_F)(V_n-V_2)].
\end{eqnarray}
Moreover, based on Eq.(4),
the current $I_n$ flowing from lead $n$ to the central region
can be obtained by summing over all the local current
$\sum_i J_{i,i+\delta y}$ at the boundary between lead $n$ and the central region.
In comparison with $I_n$ acquired from the previous calculation of $R_L$ and $R_{NL}$,
we can further testify the correctness of Eq.(4)
and the accuracy of our calculation program.

In Fig.2, we first show the spatial distributions of the local current
inside the black dashed rectangle marked in Fig.1,
when the Rashba effect is fixed at $\lambda_R=0.1$.
Throughout this work, we only discuss the negative part of the Fermi energy $E_F$ for simplicity,
and the conclusion for positive $E_F$ is the same.
From Fig.2(a) to 2(c), the Fermi energy locates at
$E_F=-0.15$, $-0.2$ and $-0.3$, respectively.
Here, the size and the orientation of the arrows indicate
the strength and the direction of the local-current flow.

As shown in Fig.2(a),
it is obvious that there exists a counter-clockwise vortex of the local current
between the two nonlocal measuring terminals\cite{foot3},
which seems like the whirlpools obtained in viscous hydrodynamic systems.
Importantly, the appearance of this vortex
induces a kind of backflow current flowing from lead 4 to lead 3,
which is in the direction opposite to the injected current.
Consequently, there exists a competition between the positive voltage drop
caused by the Ohmic transport
and the negative one caused by the backflow current.
Thus, it is natural for us to anticipate that if the ``vortex strength''\cite{foot4} is strong enough,
a negative voltage drop between $V_3$ and $V_4$ will be detected,
and a negative nonlocal resistance $R_{NL}$ will be obtained consequently
as people have shown in previous work.

\begin{figure}[h]
\includegraphics [width=\columnwidth, viewport=0 0 527 513, clip]{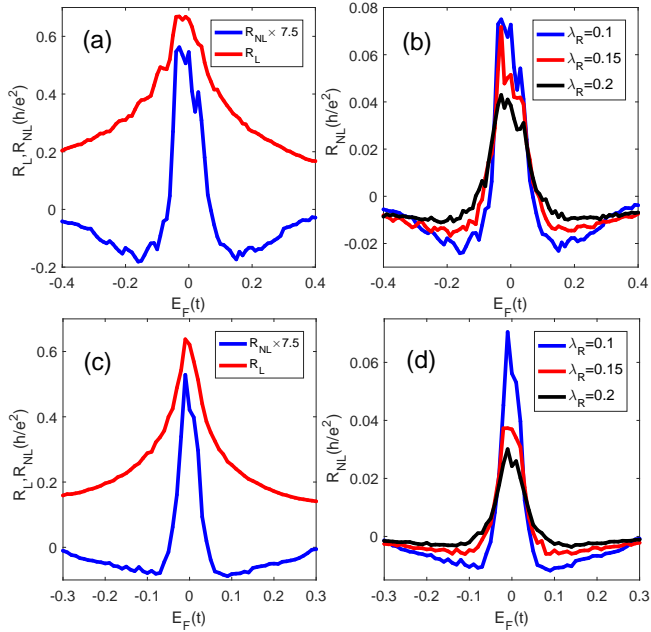}
\caption{
(a) The local resistance $R_L$ (red) and nonlocal resistance $R_{NL}$ (blue) when $\lambda_R=0.1$.
(b) The nonlocal resistance $R_{NL}$ with different Rashba strengths.
The negative peak of $R_{NL}$ reaches its maximum at $\lambda_R=0.1$ (blue),
and gradually decreases with the increase of $\lambda_R$ (red and black).
(c) and (d) The same results of panels (a) and (b) with double size parameters,
where the oscillations seem not evident now.
}\label{bulkbandsevolution}
\end{figure}

\begin{figure*}[htbp!]
\scalebox{2}{\includegraphics [width=\columnwidth, viewport=0 0 1002 435, clip]{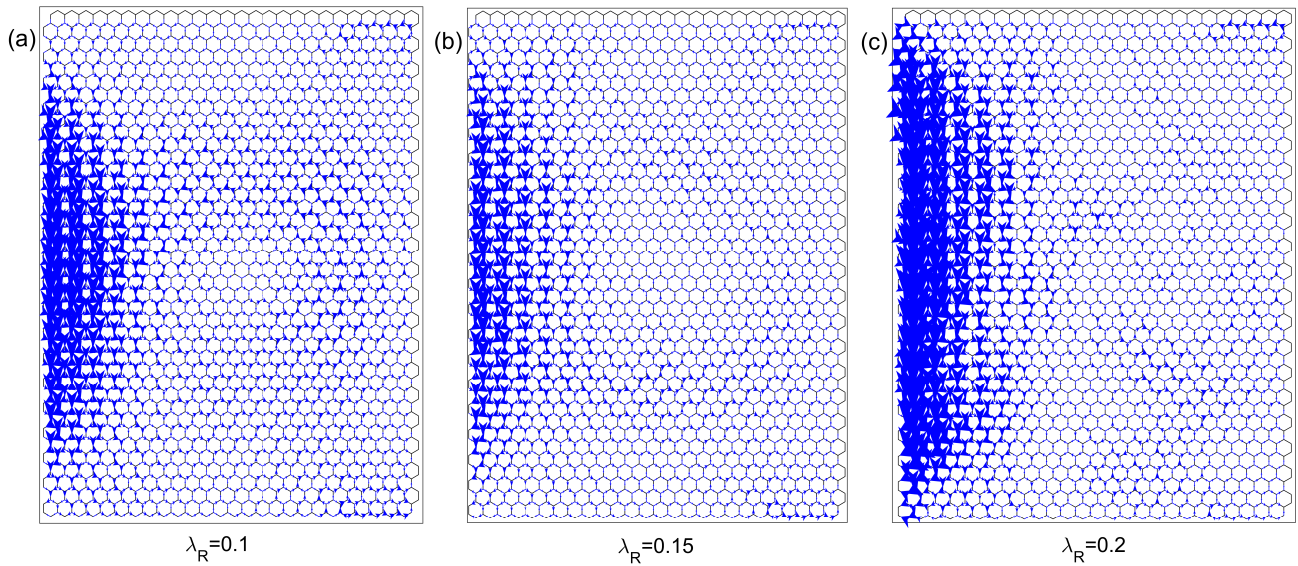}}
\caption{
The local current flow in the black dashed rectangle marked in Fig.1 when $E_F$ is fixed at -0.15.
From panel (a)-(c), the Rashba strength increases from $\lambda_R=0.1$ to 0.15 and 0.2, respectively.
The size of each arrow is proportional to the strength of the local current vector.
We find that the vortex is evident with a smaller Rashba effect,
and gradually disappears with the increase of $\lambda_R$.
Since the vortex in panel (c) is too weak,
the size of the arrows in this panel has been strengthened by two times
compared with those in panels (a) and (b).
All three panels have sufficiently high resolutions to be zoomed in to show the vortex clearly.
}\label{bulkbandsevolution}
\end{figure*}

Moreover, from Fig.2(a) to 2(c), we find that the strength of the vortex
reaches its maximum at $E_F=-0.15$,
then weakens gradually with the increase of the Fermi energy,
and finally disappears when the Fermi energy is large enough.
In order to confirm the relationship
between the vortex of the local current and the appearance of the negative $R_{NL}$,
in Fig.3(a), we show how $R_L$ and $R_{NL}$ vary with the Fermi energy $E_F$
when the Rashba effect is fixed at $\lambda_R=0.1$.
The macroscopic oscillations come from the resonance in the Fabry-P\'erot cavity,
because the size of our system is very small.
As a contrast, in Fig.3(c) and 3(d),
we show the result with a system twice the size of that in Fig.3(a) and 3(b),
where the oscillations become relatively negligible now.
Notably, in Fig.3(c) and 3(d),
we find the whole shape of $R_{NL}$ vs $E_F$ shrinks along the $E_F$ axis
and the negative peak move towards the original point.
Since the Rashba effect always appears in the form of $\lambda_R/E_F$ during the calculation,
the shrink of $E_F$ indicates that a smaller $\lambda_R$ is needed with a larger system size,
which is more practical in experiments.
In Fig.3(a), it is obvious that the blue lines of $R_{NL}$ exhibits a giant peak near the Dirac point,
then decays rapidly to its negative maximum at about $E_F=-0.15$,
and finally approaches zero as the absolute value of $E_F$ continues to increase.
The giant peak of $R_{NL}$ has been discussed in our previous work\cite{Wang},
which is assumed resulting from the extremely small density of states at the Dirac point,
and has no relationship to the vortex in our system.
Thus, in order to eliminate this effect,
we only consider the region where $E_F<-0.15$ throughout this work.
As expected, by comparing Fig.2 and Fig.3(a),
the most negative $R_{NL}$ locates exactly where the strongest vortex emerges,
and the negative value of $R_{NL}$ decreases as the vortex disappears gradually.
Thus, we can make a conclusion that the negative value of $R_{NL}$
originates from the vortex emerging between the nonlocal measuring terminals.
The stronger the vortex exhibits, the more negative $R_{NL}$ manifests.

Another method to further study the relationship between the negative value of $R_{NL}$
and the strength of the vortex
is to alter the Rashba strength $\lambda_R$ while $E_F$ remains unchanged.
In Fig.3(b), we show how $R_{NL}$ varies with $E_F$ under three different Rashba strengths
$\lambda_R=0.1$, 0.15 and 0.2.
As one can see, apart from the region close to the Dirac point,
the negative value of $R_{NL}$ decreases with increasing $\lambda_R$.
This phenomenon can be understood as follows:
if the Rashba effect is extremely strong,
the current injected into lead 1 will first transport to lead 3 along the upper edge of the system
due to the spin Hall effect,
then undergoes collisions with the boundary of the system,
and finally flows from lead 3 to lead 4.
As above, the Rashba effect actually makes a positive contribution to the nonlocal resistance $R_{NL}$,
which is in consistent with our previous calculation of $R_{Hall}$ shown in Ref.[\onlinecite{Wang}].
However, since the Rashba effect and the vortex always coexist with each other in reality,
the local-current flow induced by the Rashba effect
will eliminate the strength of the vortex caused by the ballistic transport in a certain degree.
Consequently, the vortex should gradually disappear as the Rashba effect strengthens.

Following this line of reasoning, in Fig.4, we show the spatial distributions of the local current
inside the black dashed rectangle marked in Fig.1,
when the Fermi energy is fixed at $E_F=-0.15$.
From Fig.4(a) to 4(c),
the strength of the Rashba effect increases from $\lambda_R=0.1$ to $0.15$ and $0.2$.
As expected before,
the vortex exhibits most obviously at $\lambda_R=0.1$,
and gradually disappears with the increase of $\lambda_R$.
Now, based on the calculation of the local-current flow,
we demonstrate that the ballistic transport
proposed in previous works indeed gives rise to a vortex
in the local-current distribution,
which further induces backflow current in the direction opposite to the injected
current and results in the negative $R_{NL}$ observed
both in theory and experiment.
Specifically, the stronger the vortex is, the more negative $R_{NL}$ exhibits.

\section{Two experimental criteria to the existence of the backflow current}

Interestingly, the backflow current induced by the vortex shown in Fig.2 and 4 has also been obtained
theoretically in graphene system dominated by $ee$ interaction\cite{Bandurin1,Levitov}.
To be specific, although the Hamiltonian of Eq.(1) has no terms of $ee$ interaction
and locates at the ballistic regime,
both our system and the $ee$ dominated one described by viscous hydrodynamics exhibit similar properties:
the negative value of $R_{NL}$ and the backflow current.
Significantly, in hydrodynamic theories, the vortex strength highly depends on the viscosity magnitude,
and the vortex disappears with zero viscosity.
In contrast, for our ballistic system,
according to Fig.4,
the vortex strength decreases with the increasing Rashba effect.
Thus, the backflow current and the negative value of $R_{NL}$ proposed in our ballistic system
can be controlled by tuning the Rashba strength $\lambda_R$,
which mainly relies on external electric field and can be realized in experiments.
Furthermore, since $\lambda_R$ has little relation to the viscosity,
we believe that one possible method to distinguish the appearance of the vortex current
between the ballistic and the hydrodynamic system is to tune the external electric field
in order to alter the extrinsic Rashba strength $\lambda_R$.
Specifically, by increasing the external electric field,
a reduction of the negative $R_{NL}$ will be detected in experiment,
which indicates the weakening of the backflow current in the ballistic regime.
While the hydrodynamic regime is not sensitive to the varying Rashba effect.

In addition, there also exist other signatures
for the appearance of backflow current in the ballistic regime.
For instance, the breakdown of the nonlocal WF law based on the NEGF calculation.
The local and nonlocal thermal conductances are defined as:
$\Theta_L=(T_1-T_2)/Q_1$ and $\Theta_{NL}=(T_3-T_4)/Q_1$, respectively.
Here, $T_{i}$ indicates the temperature of lead $i$
and $Q_i$ represents the heat current flowing from lead $i$ to the central region.
Based on the NEGF method, $Q_i$ can be calculated as\cite{Liu}:
\begin{eqnarray}
Q_i=\frac{1}{h}\sum_j\int{\rm d}E(E-E_F)^2f_0(1-f_0)\mathcal{T}_{ij}(E)\frac{T_i-T_j}{{k_BT}^2},
\end{eqnarray}
where $f_0(E)=1/(1+\exp\frac{E-E_F}{k_BT})$ is the Fermi distribution function.
And $\mathcal{T}_{ij}(E)$ is the transmission coefficient from lead $i$ to $j$,
which can be obtained during the calculation of $R_L$ and $R_{NL}$.

\begin{figure}[h]
\includegraphics [width=\columnwidth, viewport=0 0 560 512, clip]{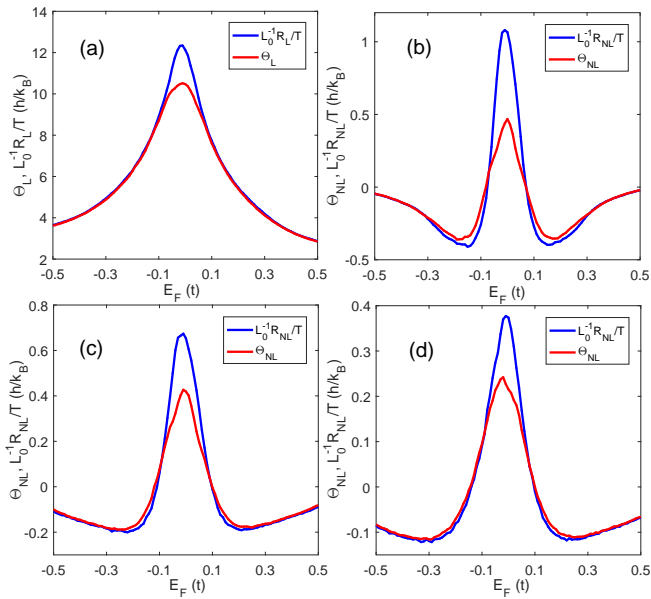}
\caption{
(a) The local thermal conductance $\Theta_L$ (red) as a function of the Fermi energy $E_F$,
compared with the local WF law $\mathcal{L}_0^{-1}R_L/T$ (blue)
at $T=500K$ and $\lambda_R=0.1$.
(b)-(d) The nonlocal thermal conductance $\Theta_{NL}$ (red)
and the corresponding nonlocal WF law $\mathcal{L}_0^{-1}R_{NL}/T$ (blue)
from $\lambda_R=0.1$ to 0.2 and 0.3, respectively.
}\label{bulkbandsevolution}
\end{figure}

In Fig.5(a), we first plot the local thermal conductance $\Theta_L$
alongside $E_F$ at $T=500K$ when $\lambda_R=0.1$.
For a direct quantitative comparison based on the WF law,
the scaled local conductance $\mathcal{L}_0^{-1}R_L/T$ in the same units as $\Theta_L$
is also shown,
where $\mathcal{L}_0=\frac{\pi^2}{3}(\frac{k_B}{e})^2$ is the Lorenz ratio.
It is obvious that these two values coincide with each other
when the absolute value of $E_F$ is larger than 0.1.
Then, in Fig.5(b),
we also show the nonlocal thermal conductance $\Theta_{NL}$
and its corresponding scaled nonlocal conductance $\mathcal{L}_0^{-1}R_{NL}/T$.
The major difference between Fig.5(a) and 5(b) is that:
the latter one exhibits an obviously additional violation of the nonlocal WF law at $-0.3<E_F<-0.1$,
while the former one does not.
Moreover, in Fig.5(b), $\mathcal{L}_0^{-1}R_{NL}/T$ is higher than $\Theta_{NL}$
near the Dirac point,
and this relationship reverses at $-0.3<E_F<-0.1$.
The above phenomena indicate that the physical pictures
behind the violations at $-0.3<E_F<-0.1$ and $-0.1<E_F<0.1$ must be totally different.
Since the vortex only exists between the nonlocal measuring terminals,
which affects the nonlocal WF law instead of the local one,
it is most likely that the violation at $-0.3<E_F<-0.1$ results from this exotic vortex.

In order to further confirm the origin of the violation at $-0.3<E_F<-0.1$,
we enlarge the Rashba strength from $\lambda_R=0.1$ to 0.2 and 0.3 in Fig.5(c) and 5(d), respectively.
It is clear that the nonlocal WF law breakdown at $-0.3<E_F<-0.1$ gradually weakens
with the increase of $\lambda_R$,
which provides evidence for the claim that this nonlocal WF law violation
actually originates from the vortex and its subsequent backflow current.
Thus, it is another feasible method in experiments that the appearance of the vortex in ballistic system
could be testified indirectly with the nonlocal thermal conductance
and the consequent breakdown of the nonlocal WF law.
More accurate analysis for the formation mechanism of this violation needs a comparison
between the local electrical current flow and local thermal current flow,
which is not shown in this work.

Above all, by altering the Rashba spin-orbital interaction strength $\lambda_R$,
two possible experimental methods have been proposed to distinguish
the appearance of the local-current vortex in ballistic regime.
The first one is to detect the negative value of $R_{NL}$ and its variation with $\lambda_R$,
and the second one is to probe the nonlocal thermal conductance $\Theta_{NL}$
and compared it with the nonlocal WF law.

\section{Discussion}

In this section, let's focus on the possible mechanisms to the formation of the exotic vortex.
First, the effect of the impurity scattering needs to be clarified in our analysis.
For comparison, we also calculate $R_{NL}$ under a weaker disorder strength $w=0.5$
compared with $w=1$ presented above.
However, the calculating results show
that $R_{NL}$ with $w=0.5$ exhibits more negative value than that with $w=1$.
Therefore, the collision between the electrons and the impurities inside
actually plays a negative role to the formation of the vortex.

\begin{figure}[h]
\includegraphics [width=5.5cm, viewport=0 0 317 414, clip]{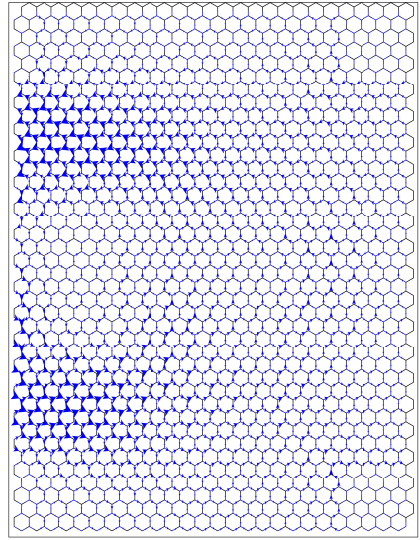}
\caption{
The local current flow in the black dashed rectangle marked in Fig.1
when a list of very strong Anderson disorder is added at the boundaries,
Other parameters are all the same as those in Fig.2(a).
It is obvious that a counter-clockwise vortex remains,
and tends to move forward into the central region compared with Fig.2(a).
}\label{bulkbandsevolution}
\end{figure}

Then, based on the Boltzmann equation,
a recently published literature proved that a vortex could also be found in the ballistic regime\cite{Chandra}.
In this semi-classical calculation,
the momentum relaxation time is set so small that the system has a sufficient long electron mean free path ($l_e$)
and definitely locates in the ballistic regime.
Thus, there must exist collisions between the flowing electrons and the boundaries,
which seems like the key role to the observation of the vortex.
In Fig.6, by adding a list of very strong Anderson disorder at the boundaries of our model,
the boundary condition of the specular reflection can be changed to a diffuse one,
which loses momentum but keeps energy conservation.
As we can see, there still exists a vortex between the nonlocal measuring terminals.
The main difference between Fig.6 and 2(a) is that
the vortex in Fig.6 tends to move left into the central region.
This phenomenon comes from the fact that the net momentum parallel to the boundary
becomes zero after the diffuse scattering,
while the net perpendicular momentum still exists.
Based on the above analysis,
we believe that the vortex directly originates from the collisions between the injected current and the boundaries,
which is the result of ballistic transport and the corresponding long $l_e$.

\begin{figure}[h]
\includegraphics [width=\columnwidth, viewport=0 0 655 376, clip]{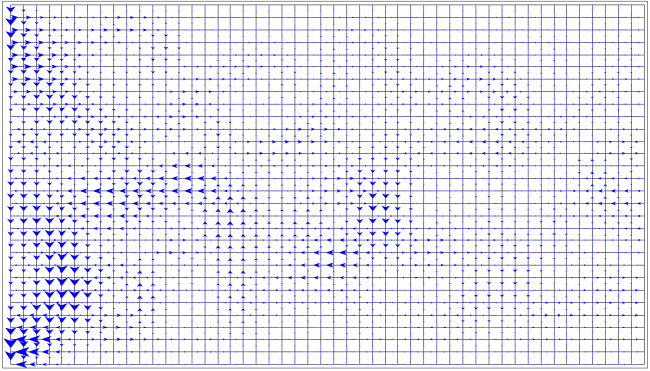}
\caption{
The local current flow of a square lattice whose size and Hamiltonian are the same as those in Fig.1.
It is obvious that no vortex exists any more.
}\label{bulkbandsevolution}
\end{figure}

Following the above analysis, the kernel question turns into
which condition could guarantee such a long $l_e$
that the system locates in the ballistic regime.
In Fig.7, we further calculate the local-current distribution of a square lattice model
whose size and Hamiltonian are the same as those of the graphene model used previously.
In order to make a better comparison between the graphene and square lattice,
the Fermi energy is fixed at $E_F=-3.8t$,
which guarantees $E_F$ locates near the bottom band of the square lattice spectrum.
Thus, both the energy and the momentum are nearly the same and consequently comparable
between the graphene and the square lattice model.

As shown in Fig.7,
though the Rashba strength is chosen as $\lambda_R=0.1$,
which is the same as that in Fig.2(a),
it is clear that only the classical Ohm-like current could be found
and no vortex exists in the square lattice model.
This phenomenon comes from the fact that at the small momentum $k$,
both the graphene and the square lattice structures have relatively large wavelength.
However, since there exists linear spectrum relationship in honeycomb lattice,
graphene is delocalized compared with the strong Anderson localization in square lattice.
Thus, $l_e$ in square lattice will be much smaller than that in honeycomb lattice.
To a certain degree,
when $l_e$ of the square lattice decreases to be shorter than the size of the system,
there is no ballistic transport,
which results in the disappearance of the vortex in square lattice as shown in Fig.7.

To sum up, we can now make a conclusion that
the vortex directly results from the elastic or semi-elastic collision between the injected current and the boundaries.
However, this collision needs the existence of a sufficient long $l_e$
and the consequent ballistic transport,
which highly relies on the linear spectrum of the graphene and the electron delocalization.
Thus, it will be much easier for us to obtain a vortex with a relatively pure graphene sample than others.

Finally, the difference between the classical Boltzmann equation and our NEGF method should also be clarified.
Specifically, the use of the classical Boltzmann equation needs a sufficient long $l_e$,
which is not calculated microscopically but only considered phenomenologically.
That is to say, the establishment of the Boltzmann equation depends on the
condition that the system has already located in the ballistic regime,
which makes no difference to the honeycomb and square lattices.
In contrast, our calculation based on the NEGF method originates from the basic Hamiltonian.
And the quantum transport could deal with different values of disorder strength,
which is more accurate and helps us better understand
the specific differences between the honeycomb and square lattice structures.

\section{Conclusion}

In conclusion, using the NEGF method,
we first calculate the local-current flow
in an H-shaped noninteracting graphene system locating in the ballistic regime.
Interestingly, an obvious vortex, similar to the one appears in a viscous hydrodynamic fluid,
can be found between the nonlocal measuring terminals.
This vortex strengthens with decreasing of the external Rashba effect,
and induces a backflow current in the direction opposite to that of the injected current.
These properties highly accord with
the confusing negative nonlocal resistance $R_{NL}$,
which has been observed in both experiments and theories previously.
The consistency comes from the fact that
the backflow current results in a negative voltage drop between the nonlocal measuring terminals,
and further induces a competition to the positive one
caused by the spin Hall transport.
Furthermore, we demonstrate that with enlarging Rashba effect strength,
both the vortex and the corresponding negative value of $R_{NL}$ become smaller,
and disappear simultaneously when the Rashba effect strength is sufficiently strong.
Thus, we conclude that the ballistic transport can also give rise to a vortex,
and may serve as a different origin of the observed negative nonlocal $R_{NL}$,
providing an alternative picture to that of the viscous hydrodynamic fluid.
We have to emphasize that the above alternative mechanism of the local current vortex does not imply
that the previous conclusion obtained from the hydrodynamics is incorrect,
but only an important supplement,
because the model and the sample size used here are quite different.

Next, we propose two experimental methods to verify the existence of vortex
in ballistic regime by tuning the strength of the external Rashba effect $\lambda_R$:
the variations of the negative value of $R_{NL}$ and the breakdown of the nonlocal WF law.
Since the $ee$ interaction dominated system is relatively insensitive to the Rashba effect,
these two methods can, in principle,
be used to distinguish the physics in the ballistic and viscous systems as well.
Moreover, since the vortex always leads to the existence of a magnetic flux,
this exotic vortex could also be detected with the newly developed
superconducting quantum interference device (SQUID)\cite{Kirtley}.

Finally and notably,
based on a comparison between our graphene system and a square lattice model,
a discussion is made that the unique linear spectrum of graphene results in a sufficient long $l_e$,
which makes the system easier to enter into the ballistic regime.
Furthermore, the exotic vortex directly originates from this ballistic transport
and the consequent collisions between the flowing current and the boundaries.

\section*{ACKNOWLEDGMENTS}

We thank the insightful discussions with Qing-feng Sun and Jie Liu.
This work was financially supported by
the Science Challenge Project (SCP) under Grant No. TZ2016003-1,
NSFC under Grants Nos. 11704348, 11534001, 11822407, 11674028,
and NBRPC under Grants Nos. 2017YFA0303301, 2017YFA0303304.


\begin{thebibliography}{10}

\bibitem{Huber}
\bibinfo{author}{M. E. Huber, N. C. Koshnick, H. Bluhm, L. J. Archuleta, T. Azua,
P. G. Bj\"{o}rnsson, B. W. Gardner, S. T. Halloran, E. A. Lucero, and K. A. Moler},
\bibinfo{journal}{Rev. Sci. Instrum}
\textbf{\bibinfo{volume}{79}}, \bibinfo{pages}{053704}
(\bibinfo{year}{2008}).

\bibitem{Abanin1}
\bibinfo{author}{D. A. Abanin, S. V. Morozov, L. A. Ponomarenko, R. V. Gorbachev, A. S. Mayorov,
M. I. Katsnelson, K. Watanabe, T. Taniguchi, K. S. Novoselov, L. S. Levitov, and A. K. Geim},
\bibinfo{journal}{Science}
\textbf{\bibinfo{volume}{332}}, \bibinfo{pages}{328}
(\bibinfo{year}{2011}).

\bibitem{Balakrishnan}
\bibinfo{author}{J. Balakrishnan, G. K. Koon, M. Jaiswal, A. H. Castro Neto, and B. \"{O}zyilmaz},
\bibinfo{journal}{Nat. Phys.}
\textbf{\bibinfo{volume}{9}}, \bibinfo{pages}{284}
(\bibinfo{year}{2013}).

\bibitem{Nowack}
\bibinfo{author}{K. C. Nowack, E. M. Spanton, M. Baenninger, M. K\"{o}nig,
J. R. Kirtley, B. Kalisky, C. Ames, P. Leubner, C. Br\"{u}ne, H. Buhmann, L. W. Molenkamp,
D. Goldhaber-Gordaon, and K. A. Moler},
\bibinfo{journal}{Nat. Mater.}
\textbf{\bibinfo{volume}{12}}, \bibinfo{pages}{787}
(\bibinfo{year}{2013}).

\bibitem{Gorbachev}
\bibinfo{author}{R. V. Gorbachev, J. C. W. Song, G. L. Yu, A. V. Kretinin, F. Withers, Y. Cao,
A. Mishchenko, I. V. Grigorieva, K. S. Novoselov, L. S. Levitov, and A. K. Geim},
\bibinfo{journal}{Science}
\textbf{\bibinfo{volume}{346}}, \bibinfo{pages}{448}
(\bibinfo{year}{2014}).

\bibitem{Shimazaki}
\bibinfo{author}{Y. Shimazaki, M. Yamamoto,	I. V. Borzenets, K. Watanabe, T. Taniguchi, and S. Tarucha},
\bibinfo{journal}{Nat. Phys.}
\textbf{\bibinfo{volume}{11}}, \bibinfo{pages}{1032-1036}
(\bibinfo{year}{2015}).

\bibitem{Sui}
\bibinfo{author}{M. Sui, G. Chen, L. Ma, W. Shan, D. Tian, K. Watanabe, T. Taniguchi, X. Jin, W. Yao,	D. Xiao, and Y. Zhang},
\bibinfo{journal}{Nat. Phys.}
\textbf{\bibinfo{volume}{11}}, \bibinfo{pages}{1027-1031}
(\bibinfo{year}{2015}).

\bibitem{Michihisa}
\bibinfo{author}{M. Yamamoto, Y. Shimazaki, I. V. Borzenets, and S. Tarucha},
\bibinfo{journal}{J. Phys. Soc. Jpn}
\textbf{\bibinfo{volume}{84}}, \bibinfo{pages}{121006}
(\bibinfo{year}{2015}).

\bibitem{Bandurin1}
\bibinfo{author}{D. A. Bandurin, I. Torre, R. Krishna Kumar, M. Ben Shalom, A. Tomadin,
A. Principi, G. H. Auton, E. Khestanova, K. S. Novoselov, I. V. Grigorieva,
L. A. Ponomarenko, A. K. Geim, and M. Polini},
\bibinfo{journal}{Science}
\textbf{\bibinfo{volume}{351}}, \bibinfo{pages}{1055}
(\bibinfo{year}{2016}).

\bibitem{Bandurin2}
\bibinfo{author}{D. A. Bandurin, A. V. Shytov, L. S. Levitov, R. K. Kumar, A. I. Berdyugin,
M. B. Shalom, I. V. Grigorieva, A. K. Geim, and G. Falkovich},
\bibinfo{journal}{Nat. Commun.}
\textbf{\bibinfo{volume}{9}}, \bibinfo{pages}{4533}
(\bibinfo{year}{2018}).

\bibitem{Braem}
\bibinfo{author}{B. A. Braem, F. M. D. Pellegrino, A. Principi, M. Roosli, C. Gold,
S. Hennel, J. V. Koski, M. Berl, W. Dietsche, W. Wegscheider, M. Polini,
T. Ihn, and K. Ensslin},
\bibinfo{journal}{Phys. Rev. B}
\textbf{\bibinfo{volume}{98}}, \bibinfo{pages}{241304(R)}
(\bibinfo{year}{2018}).

\bibitem{Wu}
\bibinfo{author}{Y. Wu, L. Zhang, C. Li, Z. Zhang, S. Liu, Z. Liao, and D. Yu},
\bibinfo{journal}{Adv. Mater.}
\textbf{\bibinfo{volume}{30}}, \bibinfo{pages}{1707547}
(\bibinfo{year}{2018}).

\bibitem{Berdyugin}
\bibinfo{author}{A. I. Berdyugin, S. G. Xu, F. M. D. Pellegrino, R. Krishna Kumar, A. Principi,
I. Toree, M. Ben Shalom, T. Taniguchi, K. Watanabe, I. V. Grigorieva, M. Polini, A. K. Geim,
and D. A. Bandurin},
\bibinfo{journal}{Science}
\textbf{\bibinfo{volume}{364}}, \bibinfo{pages}{162}
(\bibinfo{year}{2019}).

\bibitem{Levitov}
\bibinfo{author}{L. Levitov, and G. Falkovich},
\bibinfo{journal}{Nat. Phys.}
\textbf{\bibinfo{volume}{12}}, \bibinfo{pages}{672}
(\bibinfo{year}{2016}).

\bibitem{Alekseev}
\bibinfo{author}{P. S. Alekseev},
\bibinfo{journal}{Phys. Rev. Lett.}
\textbf{\bibinfo{volume}{117}}, \bibinfo{pages}{166601}
(\bibinfo{year}{2016}).

\bibitem{Levin}
\bibinfo{author}{A. D. Levin, G. M. Gusev, E. V. Levinson, Z. D. Kvon, and A. K. Bakarov},
\bibinfo{journal}{Phys. Rev. B}
\textbf{\bibinfo{volume}{97}}, \bibinfo{pages}{245308}
(\bibinfo{year}{2018}).

\bibitem{Shytov}
\bibinfo{author}{A. Shytov, J. F. Kong, G. Falkovich, and L. Levitov},
\bibinfo{journal}{Phys. Rev. Lett.}
\textbf{\bibinfo{volume}{121}}, \bibinfo{pages}{176805}
(\bibinfo{year}{2018}).

\bibitem{Wang}
\bibinfo{author}{Z. Wang, H. Liu, H. Jiang, and X. C. Xie},
\bibinfo{journal}{Phys. Rev. B}
\textbf{\bibinfo{volume}{94}}, \bibinfo{pages}{035409}
(\bibinfo{year}{2016}).

\bibitem{Tuan}
\bibinfo{author}{D. Van Tuan, J. M. Marmolejo-Tejada, X. Waintal,
B. K. Nikolic, S. O. Valenzuela and S. Roche},
\bibinfo{journal}{Phys. Rev. Lett.}
\textbf{\bibinfo{volume}{117}}, \bibinfo{pages}{176602}
(\bibinfo{year}{2016}).

\bibitem{Chandra}
\bibinfo{author}{M. Chandra, G. Kataria, D. Sahdev, and R. Sundararaman},
\bibinfo{journal}{Phys. Rev. B}
\textbf{\bibinfo{volume}{99}}, \bibinfo{pages}{165409}
(\bibinfo{year}{2019}).

\bibitem{Huang}
\bibinfo{author}{C. Huang, Y. D. Chong, and M. A. Cazalilla},
\bibinfo{journal}{Phys. Rev. Lett.}
\textbf{\bibinfo{volume}{119}}, \bibinfo{pages}{136804}
(\bibinfo{year}{2017}).

\bibitem{Gallagher}
\bibinfo{author}{P. Gallagher, C. Yang, T. Lyu, F. Tian, R. Kou, H. Zhang,
K. Watanabe, T. Taniguchi, and F. Wang},
\bibinfo{journal}{Science}
\textbf{\bibinfo{volume}{364}}, \bibinfo{pages}{158}
(\bibinfo{year}{2019}).

\bibitem{Jong}
\bibinfo{author}{M. J. M. de Jong, and L. W. Molenkamp},
\bibinfo{journal}{Phys. Rev. B}
\textbf{\bibinfo{volume}{51}}, \bibinfo{pages}{13389}
(\bibinfo{year}{1995}).

\bibitem{Scaffidi}
\bibinfo{author}{T. Scaffidi, N. Nandi, B. Schmidt, A. P. Mackenzie, and J. E. Moore},
\bibinfo{journal}{Phys. Rev. Lett.}
\textbf{\bibinfo{volume}{118}}, \bibinfo{pages}{226601}
(\bibinfo{year}{2017}).

\bibitem{Lucas}
\bibinfo{author}{A. Lucas, and K. C. Fong},
\bibinfo{journal}{J. Phys.: Condens. Matter}
\textbf{\bibinfo{volume}{30}}, \bibinfo{pages}{053001}
(\bibinfo{year}{2018}).


\bibitem{Crossno}
\bibinfo{author}{J. Crossno, J. K. Shi, K. Wang, X. Liu, A. Harzheim, A. Lucas,
S. Sachdev, P. Kim, T. Taniguchi, K. Watanabe, T. A. Ohki, and K. C. Fong},
\bibinfo{journal}{Science}
\textbf{\bibinfo{volume}{351}}, \bibinfo{pages}{1058}
(\bibinfo{year}{2016}).


\bibitem{foot1}
The calculation process is actually based on a six-terminal system.
That is to say, there exist additional two leads at the left and right side of the center region,
marked by lead L and lead R,
because the definitions of $R_L$ and $R_{NL}$ in some experiments require these two leads.
However, the deductions of $R_{L,NL}$ and $\Theta_{L,NL}$ in our paper
need no imformation from lead L and lead R,
and we just assume the both the electrical and the thermal current
flowing through lead L and lead R is zero.
Therefore, we claim that it is a four-terminal system that we use for simplicity.

\bibitem{foot2}
The $ee$ collision mean free path $l_{ee}$ is usually on the order of $100nm$ in graphene.
Thus, this system definitely lies in the ballistic regime
no matter whether the $ee$ interaction exists or not in Eq.1.

\bibitem{Jauho}
\bibinfo{author}{A.-P. Jauho, N. S. Wingreen, and Y. Meir},
\bibinfo{journal}{Phys. Rev. B}
\textbf{\bibinfo{volume}{50}}, \bibinfo{pages}{5528}
(\bibinfo{year}{1994}).

\bibitem{Jiang}
\bibinfo{author}{H. Jiang, L. Wang, Q. Sun, and X. C. Xie},
\bibinfo{journal}{Phys. Rev. B}
\textbf{\bibinfo{volume}{80}}, \bibinfo{pages}{165316}
(\bibinfo{year}{2009}).

\bibitem{book}Electronic Transport in Mesoscopic Systems, edited by S. Datta (Cambridge University Press, Cambridge, England, 1995).

\bibitem{foot3}
We strongly suggest the readers to zoom in Fig.2 and Fig.4 on a computer screen
for the convenience to read the vortex clearly.
These two figures have sufficiently high DPI.

\bibitem{foot4}
Actually, the vortex strength can not be well defined in the ballistic regime.
However, since the vortex is composite by a series of arrows arranging in a counter-clockwise line,
the vortex strength can be approximately estimated with the size of these counter-clockwise arrows.
Thus, from now on, we take the phrase ``vortex strength'' to represent the
``local-current strength in the vortex'' for simplicity.

\bibitem{Liu}
\bibinfo{author}{J. Liu, Q. Sun, and X. C. Xie},
\bibinfo{journal}{Phys. Rev. B}
\textbf{\bibinfo{volume}{81}}, \bibinfo{pages}{245323}
(\bibinfo{year}{2010}).

\bibitem{Kirtley}
\bibinfo{author}{J. R. Kirtley, M. B. Ketchen, K. G. Stawiasz, J. Z. Sun, W. J. Gallagher, S. H. Blanton, and S. J. Wind},
\bibinfo{journal}{Appl. Phys. Lett.}
\textbf{\bibinfo{volume}{66}}, \bibinfo{pages}{1138}
(\bibinfo{year}{1995}).

\end{thebibliography}
\end{document}